\documentclass[lettersize,journal]{IEEEtran}
\usepackage{amsmath,amsfonts}
\usepackage{algorithmic}
\usepackage{array}
\usepackage[caption=false,font=normalsize,labelfont=sf,textfont=sf]{subfig}
\usepackage{textcomp}
\usepackage{stfloats}
\usepackage{url}
\usepackage{verbatim}
\usepackage{graphicx}
\def\BibTeX{{\rm B\kern-.05em{\sc i\kern-.025em b}\kern-.08em
    T\kern-.1667em\lower.7ex\hbox{E}\kern-.125emX}}
\usepackage{balance}
\usepackage{cite}
\usepackage{epsfig}
\usepackage{graphicx}
\usepackage{amsmath}
\usepackage{amssymb}
\usepackage{url}

\usepackage{color,soul}
\soulregister\cite7 
\soulregister\citep7 
\soulregister\citet7 
\soulregister\ref7 
\soulregister\pageref7 

\begin{document}
\title{Fast and Light-Weight  Network for Single  Frame Structured Illumination Microscopy  Super-Resolution}
\author{\IEEEauthorblockN{Xi Cheng\IEEEauthorrefmark{1},
		Jun Li\IEEEauthorrefmark{1},
		Qiang Dai\IEEEauthorrefmark{2},
		Zhenyong Fu\IEEEauthorrefmark{1},
		and Jian Yang\IEEEauthorrefmark{1,*}} \\
	\IEEEauthorblockA{\IEEEauthorrefmark{1}PCA Lab, Key Lab of Intelligent
		Perception and Systems for High-Dimensional Information of
		\\Ministry of Education, and Jiangsu Key Lab of Image and Video
		Understanding for Social Security,\\
		School of Computer Science and Engineering,
		Nanjing University of Science and Technology, Nanjing, China}\\
	\IEEEauthorblockA{\IEEEauthorrefmark{2}
		School of Computer Science and Technology,
		Soochow University, Suzhou, China}
	\thanks{Corresponding author: Jian Yang (email:
		csjyang@njust.edu.cn).}}

\markboth{Journal of \LaTeX\ Class Files,~Vol.~18, No.~9, September~2020}%
{How to Use the IEEEtran \LaTeX \ Templates}

\maketitle

\begin{abstract}
Structured illumination microscopy (SIM) is an important super-resolution based microscopy technique that breaks the diffraction limit and enhances optical microscopy systems. With the development of biology and medical engineering, there is a high demand for real-time and robust SIM imaging under extreme low light and short exposure environments. Existing SIM techniques typically require multiple structured illumination frames to produce a high-resolution image. In this paper, we propose a single-frame
structured illumination microscopy (SF-SIM) based on deep learning. Our SF-SIM only needs one shot of a structured illumination frame and generates similar results compared with the traditional SIM systems that typically require 15 shots. In our SF-SIM, we propose a noise estimator which can effectively suppress the noise in the image and enable our method to work under the low light and short exposure environment, without the need for stacking multiple frames for non-local denoising. We also design a bandpass attention module that makes our deep network more sensitive to the change of frequency and enhances the imaging quality. Our proposed SF-SIM is almost 14 times faster than traditional SIM methods when achieving similar results. Therefore, our method is significantly valuable for the development of microbiology and medicine.
\end{abstract}

\begin{IEEEkeywords}
Super Resolution, Image Demoireing, Structured Illumination Microscopy.
\end{IEEEkeywords}

\section{Introduction}
\label{section:introduction}
Structured Illumination Microscopy (SIM) is a popular and powerful super-resolution technique for observing live cells (e.g., intracellular molecular structure, localization and interaction) to explore the internal mechanism of biology \cite{rodermund2021time} and medicine due to breaking the Abbe's diffraction limitation \cite{abbe1873contributions}. SIM often consists of two important steps: collect a series of low-resolution images (9 for 2D-SIM, 15 for 3D-SIM) using nonuniform illuminations under different angles and phases, employ reconstruction algorithms to synthesize high-resolution information from the sequentially collected images. Since it was first proposed by Heintzmannl \cite{heintzmann1999laterally} and Gustafsson \cite{gustafsson2000doubling}, there is a strong goal to fast obtain higher-quality SIM observations for the rapid development of various biological and medical processes \cite{quinn2021structural,navikas2021correlative}. However, both many sequential raw images and expensive super-resolution models heavily hinder SIM to achieve the goal. In live-cell super-resolution microscopy applications, SIM usually considers a trade-off of performance, computational cost, and the required number of raw images. 

Recently, the development of deep learning brings dividends to various computer vision tasks\cite{zhang2021aggregating,liu2021swin,cheng2020zero,cheng2021improved,liang2021swinir,zheng2021learning}. 
As an low-level computer vision task closely related to biology and optics, super-resolution microscopes also benefit from deep neural networks. A series of deep learning-based methods have also improved super-resolution SIM microscopy solutions to a certain extent \cite{jin2020deep,qiao2021evaluation,ling2020fast,zhang2021single,shah2021deep,christensen2021ml}. These methods learn a mapping from a set of low-resolution raw frame to a high-resolution SIM image through deep neural networks.
However, these methods have not explored how much the number of frames reconstructed by the SIM can be reduced. They also did not consider whether the SIM image could be reconstructed from a single original frame.

In this research, we demonstrate that using our designed network, any raw SIM frame shot at any angle or phase can be directly transformed to a high resolution result similar to which reconstructed by traditional methods using 15 frames, which significantly reduces the frequency of data acquisition and greatly reduces the synthesis time. To accomplish this task, we developed a Fast and Lightweight SIM super-resolution Network (FLSN), which has uniquely designed with a set of multi-kernel and multi-scale networks to learn images features from different receptive fields. We also designed a special noise estimation sub-network and bandpass attention modules based on Haar wavelet. These elements in our network can solve the noise problem in the original SIM frame and significantly improve the network performance especially for extreme low light and short exposure frames. This allows our deep learning based SIM super resolution method to have a faster imaging speed with lower phototoxicity which is significant to serve for the real application for biology and medical sciences.
\begin{figure*}[ht]
	\begin{center}
		\includegraphics[width=0.85\linewidth]{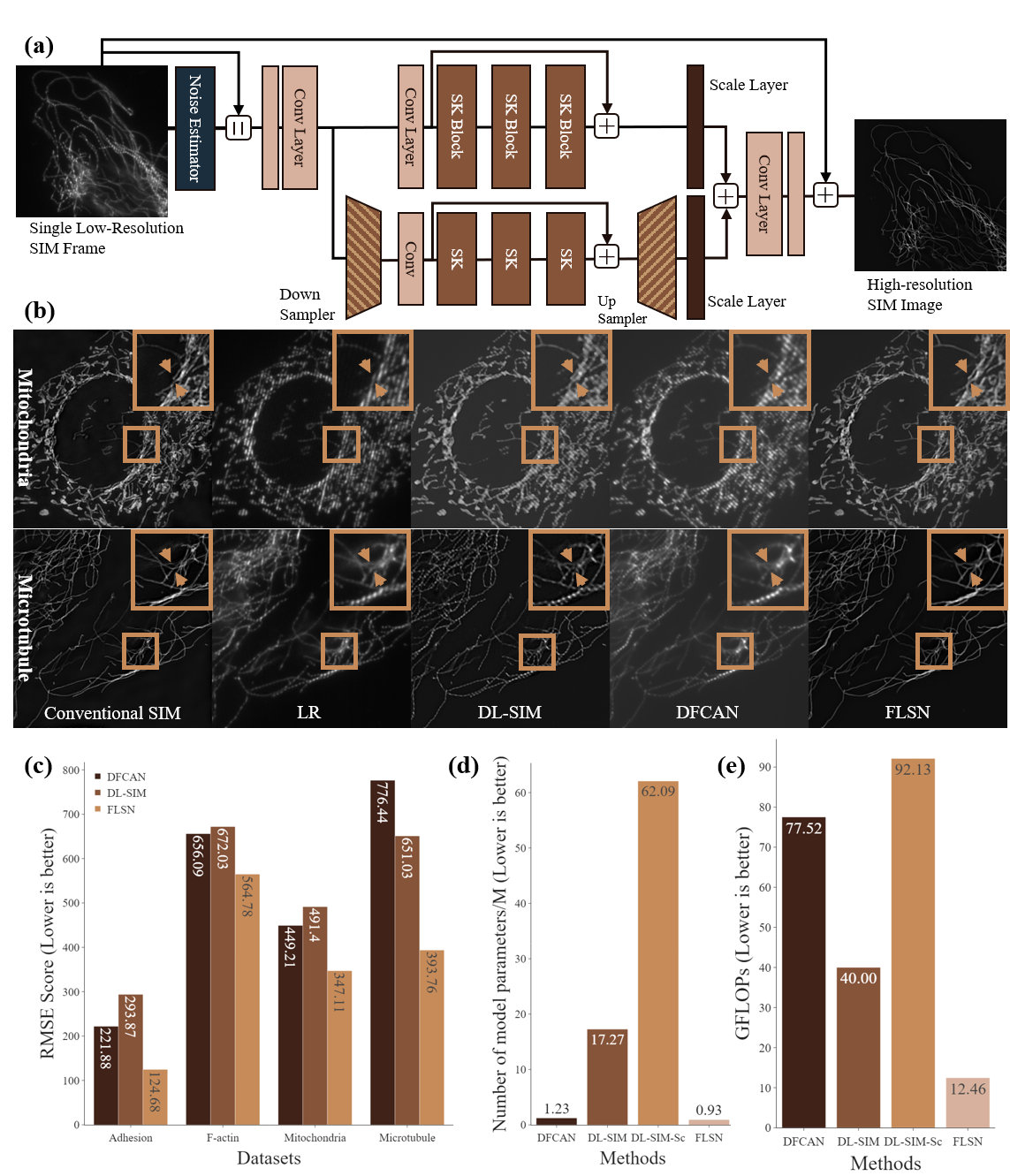}
	\end{center}
	\vspace{-0.5cm}
	\caption{\textbf{a}: The architecture of the proposed network FLSN for Single Frame SIM(SF-SIM)  \textbf{b}: Performance comparison on four data sets. \textbf{c}: Model computation(GFLOPs) comparison. \textbf{d}: Processing time comparison. \textbf{e}: Visual comparison of SF-SIM and the state of the arts.}
	\label{fig:visual_cmp_hesf}
\end{figure*}
\section{Proposed Method}
\label{section:method}
To explore the minimum number of frames for SIM reconstruction, we directly reduce the number of frames needed to one. We name the task single frame structured illumination microscopy super resolution (SF-SIM). We design a fast and light-weight multi-scale network to handle the  task and we name our neural network FLSN. 
The proposed architecture of FLSN is shown in Table~\ref{fig:visual_cmp_hesf}(a).
Our network contains three key elements: multi-scale network, noise
estimator, and bandpass attention. We used our proposed FLSN to super resolve SIM raw frame and outperform the state of the arts on 4 different datasets with lower computation and network scale.(Table~\ref{fig:visual_cmp_hesf}(c)~(e)).
The following subsections will explain the details of these elements.

\subsection{Multi-scale Network}
\label{subsection:msn}
Our proposed network contains multiple branches, which extract the features under multiple receptive fields. In the first branch, the features keep
the same scale as the input image. The height and width of the
features in the $b^{th}$ branch is $\frac{1}{2^b}$ of the first
branch. From a vertical perspective, the multi-scale network encodes
from high to low-frequency information and from short to long
dependencies in the features. This structure can effectively remove
the grating fringes and constrain moir\'{e} patterns. From a
horizontal perspective, each branch is a super-resolution sub-network
that upscales and reshapes the features to the 2$\times$ size of the
LR image. The final output HR image is the weighted combination of the
4 branches and the upscaled original frame. This is a weighted global
residual learning, jointly denoising, demoir\'{e}ing and super
resolving the LR frame. The reconstruction process can be calculated as follows:
\begin{equation}
	G_{sim}^{HR} = (\sum_{b=1}^{B}\alpha_bG^{HR}_b + I^{LR})\uparrow,
\end{equation}where $G_{sim}^{HR}$ is the final high-resolution image from
the network, $\alpha_b$ is the learnable scale factor, $G^{HR}_b$
is the high-resolution image from the $b^{th}$ branch, and $I^{LR}$ is
the input low resolution (LR) structured illumination
frame. $(\cdot)\uparrow$ indicates the 2$\times$ upscaling function.


\begin{figure}[t]
	\begin{center}
		\includegraphics[width=0.99\linewidth]{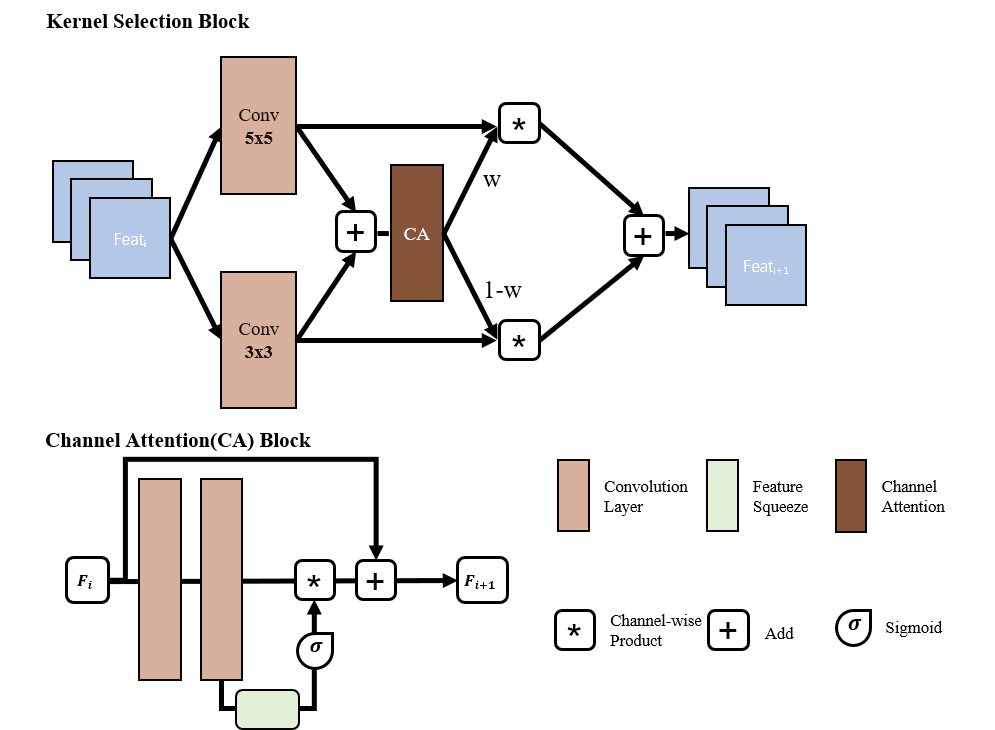}
	\end{center}
	\caption{The structure the kernel selection architecture in the branch of our backbone network.}
	\label{fig:kernel_selection}
\end{figure}

We design a basic block using kernel selection\cite{li2019selective} and channel attention\cite{hu2018squeeze} as our basic unit to construct the
subnetwork. In each branch of our network, the features are enhanced by passing
through a bandpass attention module to learn the importance of each
frequency. Then 3 basic blocks are used to
further extract the information of the enhanced features. The
architecture of our basic block is shown in
Fig.~\ref{fig:kernel_selection}. 
Different convolution kernel size can bring different receptive fields. Large receptive field can bring more semantic information while small receptive can bring local texture information\cite{ronneberger2015u}. A mixture of different kernels can enable better representation power of deep convolutional neural network. Thus, we use kernel selection\cite{li2019selective} to enhance our network backbone. This architecture contains a $3\times3$ kernel and a $5\time5$ kernel to capture different field of view. Then the features are added and passed through a channel attention to calculate the importance of different kernels. Once the weight $w$ features of $3\times3$ kernel is calculated, the features from $5\time5$ kernel is given $(1-w)$ weight. Then the weight is used to refine the features from each branch then add together as an output.
When calculating the result $Feat_{i+1}$ from the input feature $Feat_i$ with $nc$ channels, we first slice the input feature and pass them into the convolution layers with our defined kernels.
\begin{equation}
	F_i = W^{5\times5}(Feat_i^{[0,\frac{nc}{2}]}) + W^{3\times3}(Feat_i^{[\frac{nc}{2},nc]})
\end{equation}
Then we input $F_i$ into the channel attention subnetwork.
In which, we use a global average pooling with a bottleneck consisting of
two 1$\times$1 convolution layers and a ReLU activation layer to learn
the importance among the channels in the convolution layers. Then we
use the sigmoid function to map the importance value between 0 and 1
to reweight the channels. We finally add the input feature $F_i$ in
Fig.~\ref{fig:kernel_selection}(CA Block) with the reweighted features to construct
local residual learning~\cite{he2016deep,zhang2018image}. This is calculated as in the following
equation:
\begin{equation}
	w=F_i+F_i^2\sigma(W_u(A(W_d(Sq(F_i))))),
\end{equation}where $F_i^{out}$ is the output feature of this
group. $\sigma$ is the sigmoid function. $W_u$ and $W_d$ denote the
1$\times$1 convolution layers. $A(\cdot)$ means the ReLU activation
function and $Sq(\cdot)$ means the squeeze function using global
average pooling.
This helps our block learn to enhance the useful channels in the
meanwhile suppress the useless channels, leading to a selection for the channels
from $3\times3$ convolution and $5\times5$ convolution.
Finally, the output feature $Feat_{i+1}$ from our basic block is calculated as followed:
\begin{equation}
	Feat_{i+1} = W^{5\times5}(Feat_i^{[0,\frac{nc}{2}]})*w + W^{3\times3}(Feat_i^{[\frac{nc}{2},nc]})*(1-w)
\end{equation}

\subsection{Noise Estimator}
\label{subsection:ne}
\begin{figure}[t]
	\begin{center}
		\includegraphics[width=0.99\linewidth]{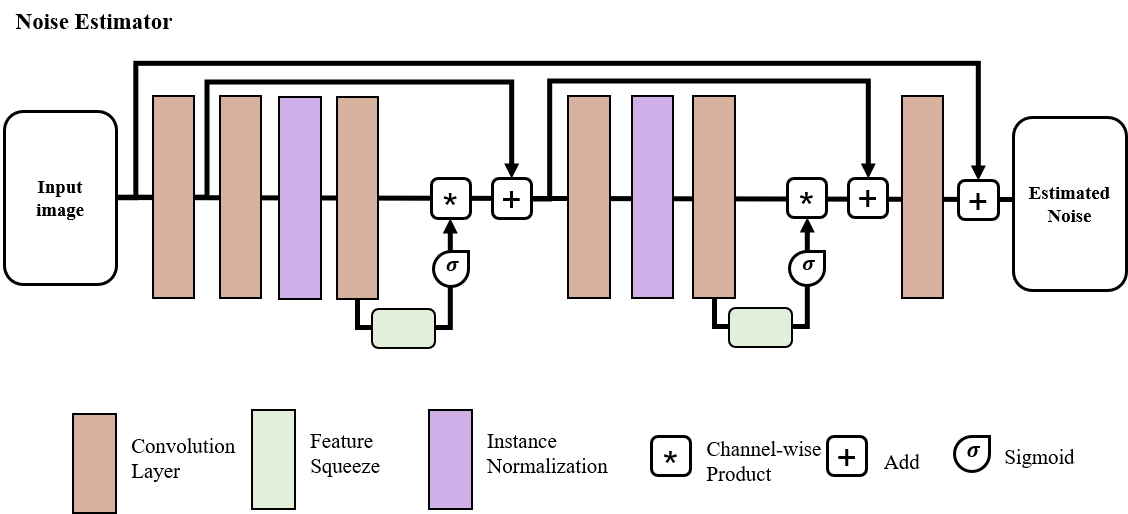}
	\end{center}
	\caption{The structure of the noise estimator module in our
		network.}
	\label{fig:noise_est}
\end{figure}

Noise is an important factor that seriously degrades the SIM imaging
quality. Usually, the noise levels in structured illumination images are
dynamic according to the exposure time and microscopy illumination
intensity. Thus, we propose a noise estimator module in the head of
the proposed network for blindly denoising~\cite{Zhang2016Beyond,guo2019toward} the input features. We
follow the idea of DnCNN~\cite{Zhang2016Beyond} to build the
module. The detailed design of our noise estimator is shown in
Fig.~\ref{fig:noise_est}. This module consists of 2 channel attention residual blocks with instance normalization~\cite{karras2019style,karras2020analyzing}. Except for the first and last layers, the convolutional layers
are followed by a ReLU activation
layer. This module helps blindly estimate the noise map for each
structured illumination frame and can be calculated as in the
following equation:
\begin{equation}
	F_{noise} = W_8(A_7(W_7(\cdots A_1(W_1(I^{LR}))))),
\end{equation}where $F_{noise}$ is the estimated noise map and
$I^{LR}$ is the low resolution structured
illumination microscopy frame. $W_i$ denotes the $i^{th}$
convolutional layer and $A_i$ is the $i^{th}$ ReLU activation layer.
The input feature of the multi-scale network is the concatenation of
the original frame and the estimated noise map:
\begin{equation}
	F_{cat} = C(I^{LR},F_{noise}),
\end{equation}where $F_{cat}$ is the combination of the upscaled LR
image and the estimated noise map. $C(\cdot)$ is the concatenation
function that stacks the two features along the channel dimension.

With the help of the proposed noise estimator, the network performance
can be significantly improved. The experimental analysis of this
module is described in the ablation study section.

\subsection{Bandpass Attention}
\label{subsection:ba}
\begin{figure}[t]
	\begin{center}
		\includegraphics[width=0.99\linewidth]{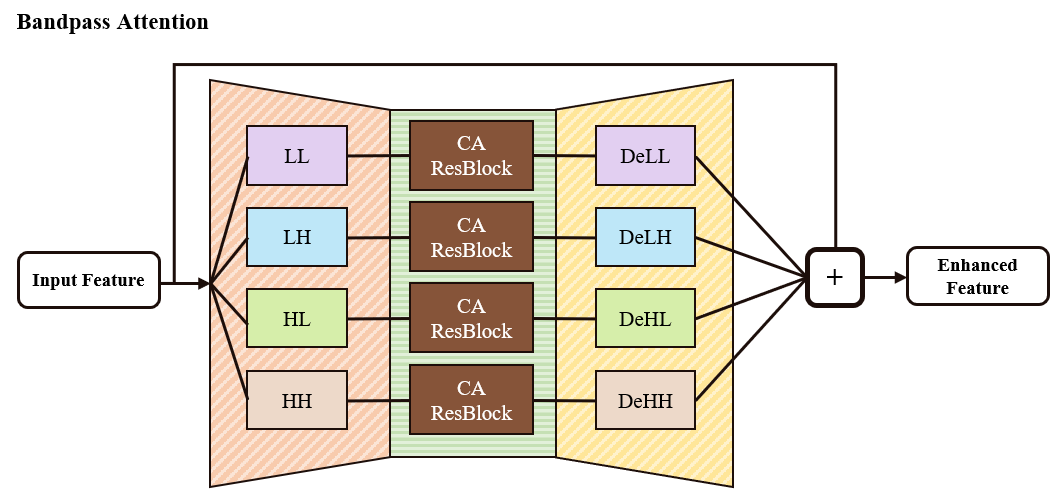}
	\end{center}
	\caption{The structure of our proposed bandpass attention which used in the full scale of the proposed FLSN network.}
	\label{fig:bp_attention}
\end{figure}

SIM imaging is based on the moir\'{e} phenomenon. Usually there are moir\'{e} and grating texture residues on the SIM frames
Moir\'{e} patterns are usually a mixture of multiple spatial
frequencies\cite{sun2018moire,cheng2021improved}. Thus separating the information and learning the
importance among different frequency bands is important.
In our work, we separate and extract the features from different frequency bands.
Our method learns a scale factor to reweight the information of the
components from different frequency bands. The module enhances the
useful frequency components and suppress the useless frequency
features. The features are mixed at the end of the module which
enhances the information in the original input. The structure 
of bandpass attention is shown in Fig.~\ref{fig:bp_attention}. We use
Haar wavelet\cite{papageorgiou1998general} to subtract features from low to high-frequency bands,
i.e. High-High (HH), High-Low (HL), Low-High (LH) and Low-Low (LL).
The Low (L) and High (H) pass filters are shown as the following
equations:
\begin{equation}
	L^T = \frac{1}{\sqrt{2}}[1\quad 1],
\end{equation}
and
\begin{equation}
	H^T = \frac{1}{\sqrt{2}}[-1\quad 1].
\end{equation}The features first input into four wavelet convolutional
layers at the first stage of the module. The input information is
downsampled and separated into 4 features from low to high
frequencies. In the second stage, each separated output is passed
through a channel attention block with residual connections (mentioned in Fig.\ref{fig:kernel_selection}. CA block) to further extract features under each
frequency band. In the third stage, we use 4 transposed convolutional
layers to upsample the information in each branch. The calculation of
each branch in the proposed module is shown as in the following formulas:
\begin{equation}
	F_{LL} = w_1*LL^T(CARB(LL(F^b_{in}))),
\end{equation}
\begin{equation}
	F_{LH} = w_2*LH^T(CARB(LH(F^b_{in}))),
\end{equation}
\begin{equation}
	F_{HL} = w_3*HL^T(CARB(HL(F^b_{in}))),
\end{equation}
\begin{equation}
	F_{HH} = w_4*HH^T(CARB(HH(F^b_{in}))),
\end{equation}where $F^b_{in}$ is the input feature from the current
branch. $w_1$ to $w_4$ mean the weight of each sub-branch of different
frequency features. $LL$ to $HH$ indicate the convolutional layers
with different frequency Haar wavelet kernels. $LL^T$ to $HH^T$
represent the transposed convolutional layers with $LL$ to $HH$ Haar
wavelet kernels. Finally, we aggregate the features in the form of
local residual learning as in the following equation.
\begin{equation}
	F_{out} = F_{LL}+F_{LH}+F_{HL}+F_{HH}+F^b_{in},
\end{equation}where $F_{out}$ is the output feature of the mixture of
reweighted information of each subbranch. The features of the
important frequencies are enhanced and the useless features are
suppressed. This forms a bandpass level attention and further enhances
the extracted features in the sub-branch. Further experimental
analysis on this module is given in ablation study section.

\subsection{Loss Function}
\label{subsection:loss}
In many image restoration tasks, $L_2$ Loss is proved to usually
provide over-smoothed results~\cite{lai2017deep}. Thus we train our
proposed SF-SIM network by optimizing $L_1$ Loss. Given a batch of
training image set $\{I^{LR}_{i,j},I^{HR}_{i}\}^{N,15}_{i=1,j=1}$, the
$L_1$ loss is calculated as in the following equation:
\begin{equation}
	L(\Theta) = \frac{1}{N}\sum_{i=1}^{N}||G_{sim}^{HR}(I^{LR}_{i,*})-I^{HR}_{i}||_1,
\end{equation}where $\Theta$ means the learnable parameter of our
network. $i$ denotes the $i^{th}$ sample in the batch and $N$ is the
total number of images in one training batch. $G_{sim}^{HR}$ denotes the
proposed SF-SIM network that generates high resolution (HR) structured
illumination microscopic image. $I^{LR}_{i,*}$ represents a random
chosen LR frame among 15 LR structured illumination (SI) frames
(i.e. 3 angles and 5 phases). $I^{HR}_{i}$ means the traditional SIM
reconstructed HR image using 15 frames shot under the regular
illumination and exposure time environment. 

\subsection{Experiment Setup\& Datasets}
\label{subsection:sexp_setup}
The detailed experimental setup is shown in
Table~\ref{table:exp_setup}. In terms of hardware, we use a server
with an NVIDIA RTX2080Ti GPU to train our SF-SIM network. 
In terms of software, we used Python 3.8 and PyTorch 1.9 on the server
with Ubuntu 16.04 system to construct and train the proposed deep
neural network. We also use CUDA11.1 and CuDNN8.0.4 to accelerate the
training speed. Besides, we also use the same software and hardware
environment to reproduce the performance of the related
state-of-the-art methods we compare below. All the codes of our 
method will be open sourced for reproducibility.

\begin{table}[htb]
	\begin{center}
		\begin{tabular}{|l|l|}
			\hline
			Hardware  & Software  \\
			\hline
			\hline
			\begin{tabular}[c]{@{}l@{}}CPU: 16 Core\\    \\ RAM:
				28GB\\    \\ GPU: RTX2080Ti \\    \\ VRAM: 11GB\end{tabular}
			& \begin{tabular}[c]{@{}l@{}}Ubuntu16.04\\    \\ CUDA11.1,
				CuDNN8.0.4\\    \\ Python3.8.3\\    \\ PyTorch1.7\end{tabular}
			\\
			\hline
		\end{tabular}
	\end{center}
	\caption{Experiment hardware and software setup.}
	\label{table:exp_setup}
\end{table}

\begin{table}[htb]
	\begin{center}
		\begin{tabular}{|l|c|c|}
			\hline
			Dataname      & Num Training & Num Testing \\
			\hline
			\hline
			Adhesion      & 669          & 99          \\
			F-actin       & 882          & 126         \\
			Microtubule   & 1007         & 167         \\
			Mitochondria & 953          & 135         \\
			\hline
		\end{tabular}
	\end{center}
	\caption{Data name and training/testing numbers of samples of the
		four microscopic images datasets.}
	\label{table:exp_dat}
\end{table}

We use the SIM dataset from \cite{jin2020deep} et al. The images in the dataset were taken with Nikon N SIM. Each sample
contains two types, the first type is standard illumination, where the
exposure time is 200ms per frame and the laser power is 70mw. The
samples in this category include 15 low-resolution SI Frames 
under standard lighting and a 2x high-resolution image synthesized by
the conventional SIM algorithm. The second category is very low light
conditions. Under this condition, only 1\% laser power is used and 
exposure time is set to ultra-short. Only 20ms is used for shooting Adhesion, F-actin, Mitochondria and Microtubule. Similarly, each sample
contains 15 low-resolution images and one 2x high-resolution image
synthesized by the conventional SIM algorithm. We name the first type of image High Exposure (HE), and the second type of image named Low Exposure
(LE). We also name the high-resolution image HR, and the
low-resolution image is named LR. The number of training and testing images is shown in Table~\ref{table:exp_dat} We also used widefield image of F-actin from the dataset provided by \cite{qiao2021evaluation} et al. which contains 20160 images for training and 1920 images for testing.

\subsection{Network Training and Hyperparameters}
\label{subsection:net_train}
When training the network, we use the Adam\cite{kingma2015adam} optimizer to minimize the
$L_1$ loss function. The batch size is set to 10 when training the
network. We normalize training and test data by dividing by the
maximum value of the data range (i.e. 65535.0). The initial learning
rate is set to 0.0001. As the training Epoch increases, we reduce the
learning rate by 12 times every 20 Epochs, for a total of 70
Epochs. It takes about 190s for training one epoch with the
above-mentioned experimental setup.


\begin{figure*}[htb]
	\begin{center}
		\includegraphics[width=0.85\linewidth]{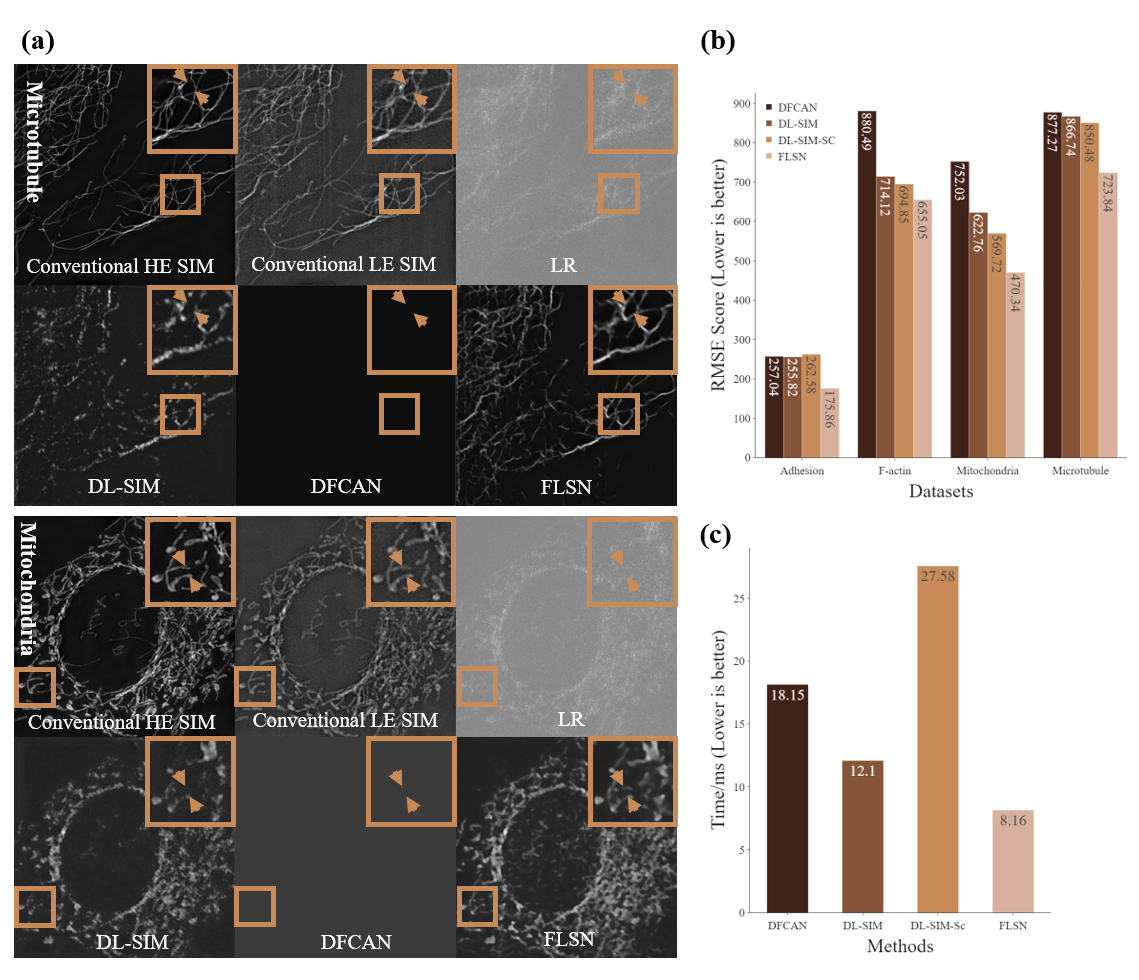}
	\end{center}
	\vspace{-0.5cm}
	\caption{\textbf{a}: Visual comparison of SF-SIM and the state of the arts under low-light and low exposure environment. \textbf{b}: Performance comparison on four data sets. \textbf{b}: Model evaluation on performance, number of parameters and computation.}
	\label{fig:visual_cmp_esf}
\end{figure*}

\section{Result}
\label{section:experiment}
\subsection{Fast Single-Frame SIM Reconstruction}
We designed a specialized network named Fast and Light-weight SIM Network (FLSN) for direct SIM reconstruction from single low-resolution SIM raw frame. Typically, traditional SIM methods need to take 9 or 15 raw frames and the recent proposed deep learning based methods\cite{jin2020deep,qiao2021evaluation} need at least 3 frames. Our proposed FLSN can reconstruct high fidelity SIM results similar to conventional methods used 15 frames.  More raw frames means more time consumed during data collection. Since just need one frame, our FLSN is very fast and light-weight which can inference at millisecond level. The multi-scale architecture and multi kernel selection in FLSN enhanced its capabilities of handling the shadow of structured illumination gratings and learn different views of features under different receptive fields. We conduct a series of experiments which illustrate the strong capability of our FLSN in different classes of cells and environments.
\begin{table}[htb]
	\centering
	\begin{tabular}{|l|c|c|c|c|}
		\hline
		\multicolumn{5}{|c|}{Single frame HE RMSE}                  \\ \hline
		Method    & Adhesion & F-actin & Mitochondria & Microtubule \\ \hline
		DFCAN     & 221.88   & 656.09  & 449.21       & 776.44      \\
		DL-SIM    & 293.87   & 672.03  & 491.4        & 651.03      \\
		FLSN Full & 124.16   & 548.23  & 331.91       & 346.96      \\ \hline
		\multicolumn{5}{|c|}{Single   frame HE SSIM}                \\ \hline
		Method    & Adhesion & F-actin & Mitochondria & Microtubule \\  \hline
		DFCAN     & 0.987    & 0.929   & 0.959        & 0.924       \\
		DL-SIM    & 0.991    & 0.928   & 0.964        & 0.960        \\
		FLSN Full & 0.995    & 0.946   & 0.977        & 0.980       \\\hline
	\end{tabular}
\vspace{0.2cm}
	\caption{RMSE and SSIM comparison on single frame SIM under normal light and  exposure environment. }
	\label{table:he}
\end{table}

Firstly, we study on the reconstruction of single raw SIM frame with normal light and exposure time (200ms). We trained and tested our FLSN using the dataset provided by Jin et al.\cite{jin2020deep} which contains 4 types of subcellular architectures including adhesion, F-actin, microtubule and mitochondria. Unlike conventional SIM methods using 15 frames(3 angles and 5 phases), FLSN can produce comparable results using only one frame. To show the strength of our proposed FLSN, we next tested and compared with the state of the art deep learning based SIM method under single raw frame settings. We evaluate the results calculating the root mean square error (RMSE) and structural similarity index(SSIM\cite{wang2004image})  (Fig.~\ref{fig:visual_cmp_hesf}(b) and \ref{table:he}) with the SIM results produced with conventional SIM algorithm. Visualization for adhension, F-actin, mitochondria and microtubule show our FLSN produce significant better SIM results than the related methods and reached comparable resolution to traditional Fourier based SIM reconstruction result(Fig.~\ref{fig:visual_cmp_hesf}(c)). Deep learning based methods are usually memory and computation consuming. The number of learnable parameters among the kernels in convolutional neural networks and giga floating-point operations per second(GFLOPs) are the two important factors that affect the network efficiency and speed. Thus we next evaluated the network scale and computation of our FLSN and the related methods(Fig. ~\ref{fig:visual_cmp_hesf}(d) and Fig.~\ref{fig:visual_cmp_hesf}(e)).

\begin{table}[htb]
	\centering
	\begin{tabular}{|l|c|c|c|c|}
		\hline
		\multicolumn{5}{|c|}{Single frame LE RMSE}                  \\ \hline
		Method    & Adhesion & F-actin & Mitochondria & Microtubule \\ \hline
		DFCAN     & 257.04   & 880.49  & 752.03       & 877.27      \\
		DL-SIM    & 255.82   & 714.12  & 622.76       & 866.74      \\
		FLSN & 167.29   & 649.27  & 461.81       & 693.21      \\ \hline
		\multicolumn{5}{|c|}{Single   frame LE SSIM}                \\ \hline
		Method    & Adhesion & F-actin & Mitochondria & Microtubule \\ \hline
		DFCAN     & 0.982    & 0.896   & 0.925        & 0.904       \\
		DL-SIM    & 0.988    & 0.922   & 0.943        & 0.920        \\
		FLSN & 0.991    & 0.93    & 0.959        & 0.938         \\\hline
	\end{tabular}
\vspace{0.2cm}
	\caption{RMSE and SSIM comparison on single frame SIM under low light and short exposure environment. }
	\label{table:le}
\end{table}

\begin{figure*}[htb]
	\begin{center}
		\includegraphics[width=0.85\linewidth]{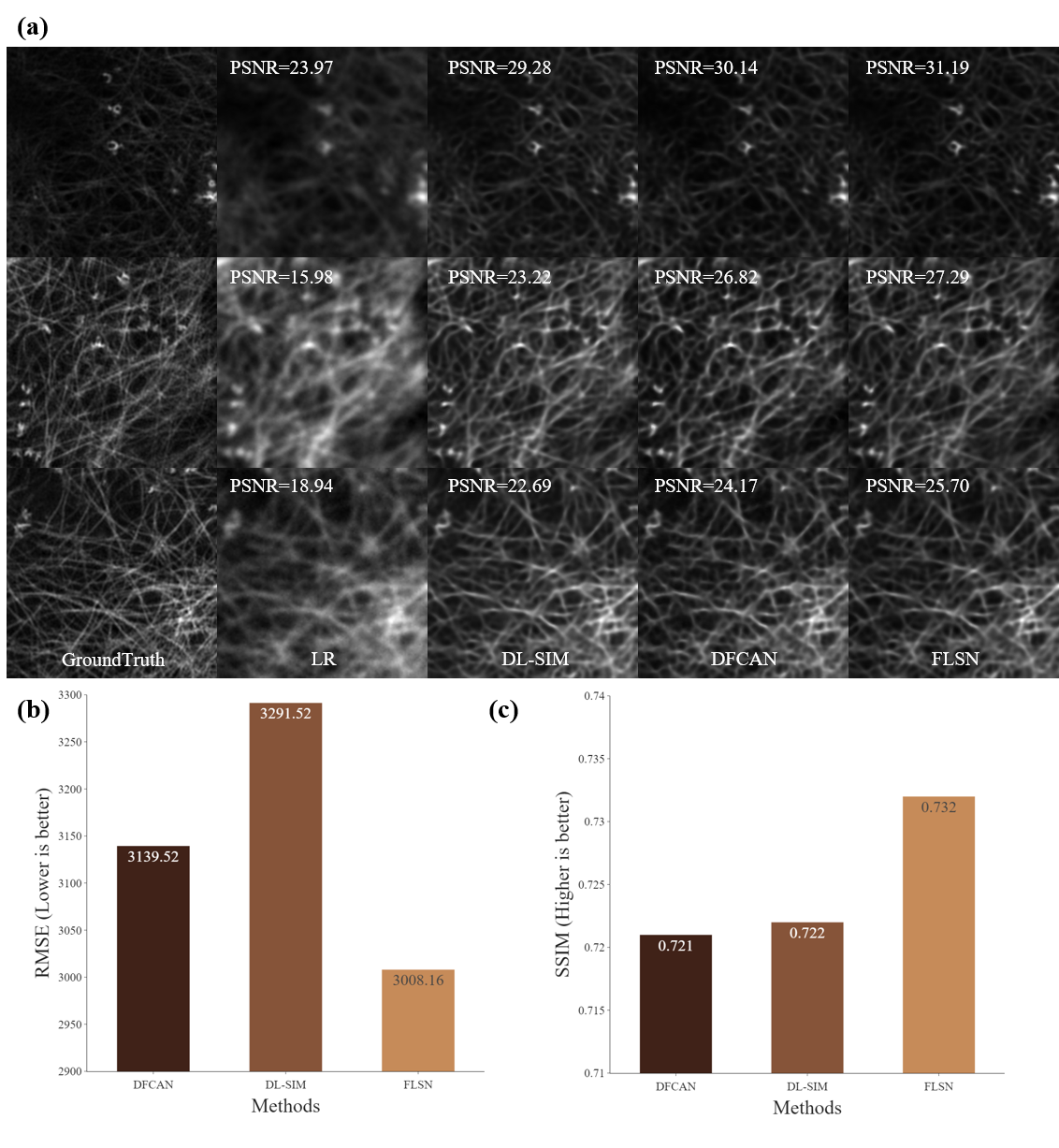}
	\end{center}
	\vspace{-0.5cm}
	\caption{\textbf{a}: Visual comparison of SF-SIM and the state of the arts under widefield environment. \textbf{b}: Performance comparison on four data sets. \textbf{c}: Model evaluation on performance, number of parameters and computation.}
	\label{fig:visual_cmp_wfsf}
\end{figure*}

\subsection{Fast SIM Reconstruction under Extreme Environment}
Our FLSN is fast and light-weight which can produce 1024$\times$1024 high resolution results within 10ms. However the slow acquisition speed for raw images strictly limits the execution time of the whole system. Thus we further train and evaluate our FLSN with image obtained under low light level and short exposure time\cite{jin2020deep} which can be obtained much faster than the normal environment(5-20ms). This type of data is much more challenging for reconstructing a high fidelity SIM result. With the decrease of the intensity of laser, content information in the raw frame become less and short exposure time result in serious noise on the image. When designing the network, we specially construct a noise estimation(NE) module. When input low exposure raw frame, NE first calculate the noise map one the image and a skip connection concatenate the raw image with the noise map. This helps our network perform significantly better under low light and short exposure environment\ref{table:le}. We also trained and tested the most related deep learning SIM methods\cite{jin2020deep,qiao2021evaluation} under this condition(Fig.\ref{fig:visual_cmp_esf}(b) and (c)).
\subsection{Fast Reconstruction for Widefield Images}
For widefield microscopic images, our proposed FLSN can still work well. Widefield images are usually more blurry and have no structured illumination pattern compared with raw SIM frames . FLSN and the related methods\cite{jin2020deep,qiao2021evaluation} were trained and tested using the widefield F-actin dataset provide by qiao et al.\cite{qiao2021evaluation}.  Similarly, RMSE(Fig.~\ref{fig:visual_cmp_wfsf}(b)) and SSIM (Fig.~\ref{fig:visual_cmp_wfsf}(c) and Table.~\ref{table:wf}) were calculated to evaluate the performance of these methods. Next, zoomed in details of the results(Fig.~\ref{fig:visual_cmp_wfsf}(a)) of these methods were compared, our FLSN still reached a comparable resolution with the conventional SIM results using only one widefield input.

\begin{table}[htb]
	\centering
	\begin{tabular}{|l|c|c|}
		\hline
		\multicolumn{3}{|c|}{Comparison on Widefield image super-resolution} \\ \hline
		Method       & Widefield F-actin RMSE    & Widefield F-actin SSIM    \\ \hline
		DFCAN        & 3139.52                   & 0.721                     \\
		DL-SIM       & 3291.52                   & 0.722                     \\
		FLSN    & 3008.80                    & 0.732                     \\\hline
	\end{tabular}
\vspace{0.2cm}
	\caption{RMSE and SSIM comparison on widefield image super resolution.}
	\label{table:wf}
\end{table}

\section{Ablation Study}
\label{section:abla}
In order to study the efficiency of our proposed modules, we conduct
ablation studies on the noise estimator
module sec.~\ref{subsection:abla_ne} and the bandpass attention
module sec.~\ref{subsection:abla_bp}.

\subsection{Noise Estimator}
\label{subsection:abla_ne}
In this section, we study the importance of the noise estimator (NE)
module in the head of our network. The noise estimator filter blindly
estimates the noise map in the original noisy image and extracts the
useful information. In Table~\ref{table:abla_dhead}, experimental
results show the network with NE reconstruct high-resolution results
with lower RMSE on both HE and LE environments on all of the four
datasets. Especially in the LE environment, the extremely low light
and shot exposure make the noise intensity significantly stronger than
the frame shot under the HE environment. The network with NE module
decreases more RMSE in the LE environment than in the HE environment,
e.g. RMSE decreases 19.07 under HE environment while decreases 88.16
under LE environment in the Adhesion dataset. Thus experimental
results support that NE is very effective and helps the network gain
higher fidelity.
\subsection{Bandpass Attention}
\label{subsection:abla_bp}
In this section, we conduct experiments to show the effectiveness of
the proposed bandpass attention (BA) module. Moir\'{e} patterns
usually consist of a large variety of frequency bands. Therefore, in
theory, BA has a strong target for structurd illumination microscopy
(SIM) image processing which is based on moir\'{e} generation. 
Table~\ref{table:bp_attention} show the RMSE score comparison on four
evaluation datasets with both HE and LE environments. As shown in the
table, among all the four datasets, the network with the BA module
performs achieves the lowest RMSE. The RMSE scores decrease 17.84 under HE environment and decrease 30.25 on adhesion under the LE
environment. Therefore, experimental results indicate that our
proposed BA module is useful for improving the network performance
when super resolving SIM images.
\begin{table*}[htb]
	\begin{center}
		\begin{tabular}{|l|c|c|c|c|}
			\hline
			Method        & Adhesion  & F-actin  & Mitochondria  & Microtubule  \\
			\hline
			\hline
			HE Without NE &  143.23  &  606.23  & 391.72  &  458.68   \\
			HE With NE    &  124.16	&	548.23  & 331.91  &  346.96  \\
			\hline
			\hline
			LE Without NE &  255.45  &  874.14 & 749.40 & 882.82 \\
			LE With NE    & 167.29	&   649.27	&  461.81  &693.21	 \\
			\hline
		\end{tabular}
	\end{center}
	\caption{Structured illumination microscopic (SIM) image super
		resolution performance comparison with/without noise estimator
		module under HE and LE environments.}
	\label{table:abla_dhead}
\end{table*}

\begin{table*}[htb]
	\begin{center}
		\begin{tabular}{|l|c|c|c|c|}
			\hline
			Method        & Adhesion  & F-actin  & Mitochondria  & Microtubule  \\
			\hline
			\hline
			HE Without BA &  142.00  &  615.59  & 373.02  &  449.57   \\
			HE With BA    &  124.16	&	548.23  & 331.91  &  346.96  \\
			\hline
			\hline
			LE Without BA &  197.54  & 672.55  & 481.45 & 751.97\\
			LE With BA    & 167.29	&   649.27	&  461.81  &693.21	 \\
			\hline
		\end{tabular}
	\end{center}
	\caption{Structured illumination microscopic (SIM) image super
		resolution performance comparison with/without our proposed
		bandpass attention module under HE and LE environments.}
	\label{table:bp_attention}
\end{table*}

\section{Discussion}
Usually conventional SIM algorithms use 15 frames(3 angles and 5 phases) for reconstruction. To enhance the speed of SIM imaging, decreasing the number of shots is necessary. There is a trade-off that multiple shot of raw frames will cost a lot of time although more frames will reconstruct better results. With the help of powerful neural networks, using fewer shots of raw frames to produce high quality SIM results become possible\cite{jin2020deep,ling2020fast}. Compared with the most related methods\cite{jin2020deep,qiao2021evaluation}, in the normal light and exposure time environment, our proposed FLSN can still use one raw frame to beat those deep learning based methods using 15 frames. Our FLSN will be much stronger and got much better results when using 15 frames.\ref{table:discussion} In the low light and short exposure environments, very little useful information is kept in the frame. In this extreme environment, FLSN with single frame input can still achieve comparable results compared with those methods using 15 frames. Also, when increasing the number of raw frames to 15, FLSN can still achieve the best results\ref{table:discussion}.
\begin{table}[htb]
	\centering
	\begin{tabular}{|l|c|c|c|c|}
		\hline
		\multicolumn{5}{|c|}{ RMSE for HE   Environment}                                    \\ \hline
		Method                  & Adhesion & F-actin & Mitochondria & Microtubule \\\hline
		DFCAN-15       & 190.72   & 614.17  & 401.28       & 765.43      \\
		DL-SIM-15       & 145.49   & 581.69  & 339.92       & 392.59      \\
		DL-SIM-SC-15  & 133.98   & 583.49  & 363.09       & 341.6       \\
		FLSN -15       & 117.32   & 561.77  & 313.08       & 267.38      \\
		FLSN -1           & 124.16   & 548.23  & 331.91       & 346.96      \\ \hline
		\multicolumn{5}{|c|}{RMSE for LE Envrionment}                                      \\ \hline
		Method                  & Adhesion & F-actin & Mitochondria & Microtubule \\\hline
		DFCAN -15        & 257.73   & 880.07  & 888.62       & 752.37      \\
		DL-SIM -15       & 170.37   & 640.49  & 434.76       & 668.33      \\
		DL-SIM-SC-15     & 157.12   & 610.67  & 380.11       & 494.43      \\
		FLSN -15      & 135.12   & 572.65  & 350.14       & 478.13      \\
		FLSN - 1        & 167.29   & 649.27  & 461.81       & 693.21    \\\hline 
	\end{tabular}
\vspace{0.1cm}
	\caption{RMSE comparison for deep learning SR SIM models with 15 frames v.s. 1 frame.}
	\label{table:discussion}
\end{table}

\section{Conclusion}
\label{section:conclusion}
Structured illumination microscopy is an important type of
super-resolution microscopy that breaks the diffraction limitation and
improved the resolution of the microscopic object under optical
microscopy systems. With the development of biology and medical
engineering, there is a high demand to improve the SIM imaging speed
and performance under extreme low-light and short exposure time
conditions. Therefore, we propose a single frame structured
illumination microscopy (SF-SIM) method based on deep learning and
convolutional neural networks and we name it FLSN. Compared with the existing SR-SIM
methods\cite{christensen2021ml,jin2020deep,ling2020fast,qiao2021evaluation,shah2021deep}, only one shot of a structured illumination frame is needed to
generate similar results as traditional SIM using 15 shots. The
proposed noise estimator and bandpass attention further enhance the
imaging quality and make the results robust under serious working
conditions. The proposed SF-SIM greatly improves the imaging speed and
enhances the super-resolution quality which is of great value for the
development of microbiology and medicine.

\section{Acknowledgment}
The authors would like to thank the editor and the anonymous reviewers
for their critical and constructive comments and suggestions.
This work was supported by the National Science Fund of China under Grant
No. U1713208, Program for Changjiang Scholars.


\bibliographystyle{IEEEtran}
\bibliography{ref}{}
\end{document}